\documentclass[english,aps,manuscript]{revtex4}
\usepackage[T1]{fontenc}
\usepackage[latin9]{inputenc}
\usepackage[active]{srcltx}
\usepackage{color}
\usepackage{units}
\usepackage{textcomp}
\usepackage{amstext}
\usepackage{amssymb}
\usepackage{graphicx}
\usepackage{subscript}

\makeatletter

\DeclareRobustCommand{\greektext}{%
  \fontencoding{LGR}\selectfont\def\encodingdefault{LGR}}
\DeclareRobustCommand{\textgreek}[1]{\leavevmode{\greektext #1}}
\DeclareFontEncoding{LGR}{}{}
\DeclareTextSymbol{\~}{LGR}{126}

\@ifundefined{textcolor}{}
{%
 \definecolor{BLACK}{gray}{0}
 \definecolor{WHITE}{gray}{1}
 \definecolor{RED}{rgb}{1,0,0}
 \definecolor{GREEN}{rgb}{0,1,0}
 \definecolor{BLUE}{rgb}{0,0,1}
 \definecolor{CYAN}{cmyk}{1,0,0,0}
 \definecolor{MAGENTA}{cmyk}{0,1,0,0}
 \definecolor{YELLOW}{cmyk}{0,0,1,0}
}

\usepackage{epstopdf}

\makeatother

\usepackage{babel}
\begin{document}
\begin{flushright}
Version: \today
\par\end{flushright}

\title{Giant Secondary Grain Growth in Cu Films on Sapphire}

\author{David L. Miller}

\email{david.miller@nist.gov}

\author{Mark W. Keller}

\email{mark.keller@nist.gov}

\author{Justin M. Shaw}

\author{Katherine~P.~Rice}

\author{Robert R. Keller}

\author{Kyle M. Diederichsen}

\address{National Institute of Standards and Technology, Boulder, CO 80305}
\begin{abstract}
{\normalsize Single crystal metal films on insulating substrates are
attractive for microelectronics and other applications, but they are
difficult to achieve on macroscopic length scales. The conventional
approach to obtaining such films is epitaxial growth at high temperature
using slow deposition in ultrahigh vacuum conditions. Here we describe
a different approach: sputter deposition at modest temperatures followed
by annealing to induce secondary grain growth. We show that polycrystalline
as-deposited Cu on $\alpha$-Al\textsubscript{2}O\textsubscript{3}(0001)
can be transformed into Cu(111) with centimeter-sized grains. Employing
optical microscopy, x-ray diffraction, and electron backscatter diffraction
to characterize the films before and after annealing, we find a particular
as-deposited grain structure that promotes the growth of giant grains
upon annealing. To demonstrate one potential application of such films,
we grow graphene by chemical vapor deposition on wafers of annealed
Cu and obtain epitaxial graphene grains of 0.2~mm diameter.}{\normalsize \par}
\end{abstract}

\thanks{Official contribution of the National Institute of Standards and
Technology; not subject to copyright in the United States.}

\maketitle
Vapor-phase thin film deposition, first enabled by the invention of
modern vacuum technology about a century ago \cite{Mattox:2003uq},
has developed into an essential component of many modern technologies.
Advances in electron diffraction and microscopy over the past century
provided tools for probing the crystalline structure of thin films
and spurred investigation of the fundamental processes that determine
grain size and orientation \cite{Pashley:1975kx}. Thus grain growth
and epitaxy have become central topics in thin film research. In applications
of thin films, much effort is spent manipulating grain structure to
achieve specific mechanical, electrical, magnetic, optical, and chemical
properties. Examples include magnetic recording media \cite{Piramanayagam:2007fk},
electrochemical catalysts \cite{Vliet:2012ys}, and interconnects
for microelectronics whose resistance to electromigration depends
on film texture and grain size \cite{Tan:2007uq}. A common challenge
for metals and semiconductors grown on insulating substrates is to
obtain a desired orientation of grains (such as <111> perpendicular
to the film plane) along with macroscopic grain sizes ($\gtrsim\unit[100]{\mu m}$
in the plane of the film). The conventional approach to this challenge
to is to seek conditions that produce large epitaxial grains during
growth. An alternative approach, and the one that is the focus of
this work, is to deposit a polycrystalline film and then produce large
epitaxial grains by annealing. Although the basic idea was clearly
described almost 25 years ago by C. V. Thompson et al., who called
it ``epitaxial grain growth'' \cite{Thompson:1990kx}, it has remained
relatively unexplored since then. Here we show that this approach
can produce films with individual grains with area $>\unit[1]{cm^{2}}$.

The work described here was motivated by a desire to produce Cu(111)
films that approach the ideal of a single crystal and that are suitable
as substrates for chemical vapor deposition (CVD) of graphene and
hexagonal boron nitride (h-BN) at temperatures near $\unit[1000]{^{\circ}C}$.
Copper is the most commonly used substrate for graphene CVD because
its negligible carbon solubility enables growth of precisely one layer
over a wide range of growth conditions. It can also be readily etched
away to allow transfer of the graphene to other substrates for further
device fabrication steps \cite{Suk:2011vn}. Although polycrystalline
Cu foils are typically used, the Cu(111) surface provides a hexagonal
template with a relatively small lattice mismatch (3.8\% and 2.2\%
for graphene and h-BN, respectively) that allows epitaxial growth
with low, uniform strain \cite{He:2012fk}. Previous work on bulk
single crystals has shown that graphene has less rotational disorder
when grown on Cu(111) than on Cu(100) \cite{Nie:2011ys,Wofford:2010fk}.
Exploiting the benefits of Cu(111) for commercial, wafer-scale production
of graphene will require a process for producing crystalline films
on a suitable substrate. A particularly attractive substrate is $\alpha$-Al\textsubscript{2}O\textsubscript{3}(0001).
It is widely used by manufacturers of radio-frequency electronics
and light-emitting diodes in the form of wafers with diameters up
to 200~mm, it is physically and chemically stable under graphene
CVD conditions, and it can likely be reused after metal etching to
release the graphene layer.

There is a large body of work pertaining to Cu on $\alpha$-Al\textsubscript{2}O\textsubscript{3},
which is a model system for epitaxy, adhesion, and other properties
of metal-ceramic interfaces \cite{Dehm:2005fk}. On the $\alpha$-Al\textsubscript{2}O\textsubscript{3}(0001)
surface, Cu grows epitaxially with a (111) texture, i.e. \textcolor{black}{$(111)_{\textrm{Cu}}$||$(0001)_{\textrm{Al}{}_{2}\textrm{O}_{3}}$,
and with} two distinct in-plane orientation relationships (ORs). The
most commonly observed OR (typically referred to as OR~I) is \textcolor{black}{$\langle110\rangle{}_{\textrm{Cu}}$||$\langle10\bar{1}0\rangle_{\textrm{Al}{}_{2}\textrm{O}_{3}}$}.
The second OR, called OR~II, is \textcolor{black}{$\langle110\rangle{}_{\textrm{Cu}}$||$\langle2\bar{1}\bar{1}0\rangle_{\textrm{Al}{}_{2}\textrm{O}_{3}}$}.
The two ORs differ by a rotation of $30^{\circ}$ about <111>. Several
studies have found that the dominant OR depends sensitively on substrate
preparation \cite{Scheu:2006bh,Oh:2006zr,Oh:2007fk}, substrate temperature
\cite{Dehm:2005fk}, deposition rate \cite{Dehm:2005fk}, and other
conditions \cite{Curiotto:2011zr}. We show here that a particular
mixture of OR~I and OR~II in as-deposited Cu films enables giant
grain growth upon annealing.

Grain growth in thin films can occur during deposition and during
subsequent processing steps such as annealing \cite{Thompson:2000fr}.
The equilibrium state of a film is determined by the interplay of
various energies: film-substrate interface, film free surface, grain
boundary, and strain. The actual state of a film is also affected
by kinetic processes such as diffusion, thus substrate temperature
and the energy of arriving species during deposition are important
in determining grain size and orientation. Due to the interplay among
various energies, conventional grain growth stagnates when the average
grain size is about $3\times$ the film thickness \cite{Thompson:2000fr}.
In some cases, particularly for Cu and other fcc metals where grain
boundaries are mobile at relatively low temperatures, grains can grow
larger than the stagnation limit through a process known as secondary
grain growth \cite{Thompson:1985mz}. This occurs when grains of a
particular low energy orientation grow at the expense of a matrix
of other, stagnated grains. Although secondary grain growth can occur
during deposition, it is typically exploited during subsequent annealing
\cite{Thompson:1990kx}, as we do here. In Cu films sputtered onto
amorphous SiO\textsubscript{2}, secondary grain growth driven by
minimization of strain energy has produced grains $\sim\unit[10]{\mu m}$
to $\sim\unit[100]{\mu m}$ across with a (100) orientation \cite{Takewaki:1995fk,Vanstreels:2008fj}.
In our epitaxial Cu films, the fact that a (111) orientation is favored
indicates surface and interface energies dominate over strain energies
\cite{Thompson:2000fr,Deng:2012kx}.

Annealing can cause a thin film to break into discontinuous islands,
a phenomenon known as dewetting or agglomeration \cite{Thompson:2012fk,Srolovitz:1995kx}.
The process, which is driven by minimization of the total energy of
the film/substrate system, typically begins with the development of
thermal grooves at grain boundaries in the film \cite{Mullins:1957fk}.
These grooves deepen at a rate that increases with temperature, and
dewetting occurs when the grooves reach the substrate, so thinner
films become discontinuous at lower temperatures. For films exposed
to temperatures near the bulk melting point, as is the case when Cu
(melting point of $\unit[1084]{^{\circ}C}$) is used for CVD of graphene,
dewetting is a major limitation on the survivability of the film.
Fortunately, grain growth and dewetting are competing processes \cite{Thompson:2000fr}.
As we show here, the growth of giant grains at temperatures somewhat
below $\unit[1000]{^{\circ}C}$ allows Cu films to endure subsequent
graphene growth conditions with minimal dewetting.

In this paper, we present an extensive investigation of grain size
and orientation for Cu films sputtered onto $\alpha$-Al\textsubscript{2}O\textsubscript{3}(0001)
substrates at a wide range of temperatures, and then annealed at temperatures
near $\unit[1000]{^{\circ}C}$. We characterize our films using x-ray
diffraction (XRD) and electron backscatter diffraction (EBSD) to measure
crystallinity, and optical microscopy to show film properties over
large areas. We also report results for large domain graphene growth
by CVD on these films, characterized by optical microscopy and Raman
spectroscopy. Our findings show that an appropriate combination of
deposition and annealing temperatures can produce Cu(111) single crystal
films, free of twinning and thermal grooves, over macroscopic length
scales. Such films offer an ideal substrate for epitaxial CVD of graphene,
h-BN, and possibly other materials.

\begin{figure}[!th]
\centering{}\includegraphics[scale=0.4]{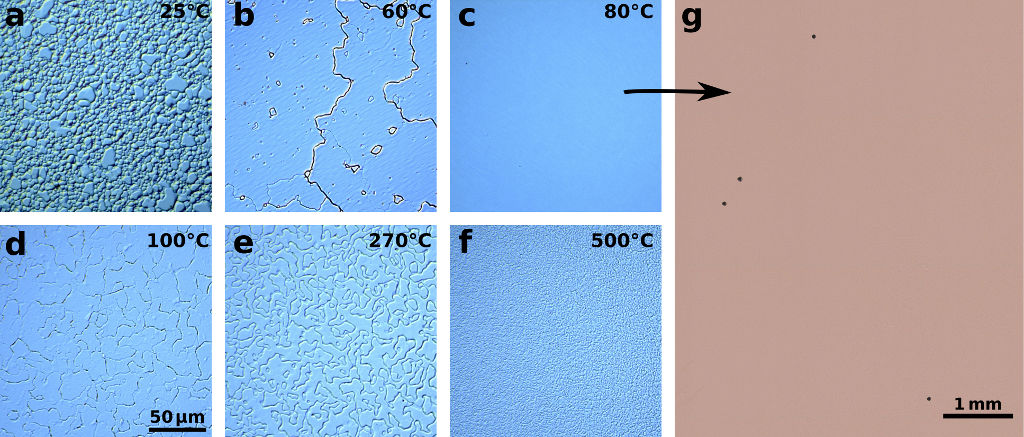}\caption{\textbf{Giant grains in Cu films after annealing at $\unit[950]{^{\circ}C}$.
a}-\textbf{f}, Optical images taken using differential interference
contrast with a $50\times$ objective lens. The scale bar in \textbf{d}
applies to images \textbf{a}-\textbf{f}. Film deposition temperature
$T_{\textrm{d}}$ is indicated in the upper right corner of each image.
Dark lines in \textbf{b},\textbf{ d}, and \textbf{e} are thermal grooves
that mark the edges of grain boundaries in the film. The region shown
in \textbf{c} is a single Cu grain.\textbf{ g}, Conventional optical
image of a larger area for $T_{\textrm{d}}=\unit[80]{^{\circ}C}$,
stitched from several smaller images, showing the absence of grain
boundaries over macroscopic areas. Dark spots are areas of dewetting.}
\label{Optical}
\end{figure}

Figure \ref{Optical} shows optical microscopy images of Cu films
deposited at various temperatures $T_{\textrm{d}}$ and annealed at
temperature $T_{\textrm{a}}=\unit[950]{^{\circ}C}$ for 40~min. The
annealing conditions were chosen to be similar to those used for graphene
CVD, but without the hydrocarbon precursor (see Methods). Images for
additional values of $T_{\textrm{d}}$ and $T_{\textrm{a}}$ are shown
in Supplementary Fig. S1. The dark lines apparent in Fig. \ref{Optical}\textbf{b},\textbf{d},\textbf{e}
are thermal grooves that mark the edges of grain boundaries in the
Cu film (see \cite{Miller:2012fk} and Supplementary Fig. S2 for confirmation
of this correspondence). Since these films are 450~nm thick, conventional
grain growth stagnation would limit grains to roughly $3\times$ the
film thickness, or $\unit[1.5]{\mu m}$. The actual grains are much
larger than this limit, and for $T_{\textrm{d}}=\unit[80]{^{\circ}C}$
the grains exceed the image size. Figure \ref{Optical}\textbf{g}
shows a larger area from a $T_{\textrm{d}}=\unit[80]{^{\circ}C}$
wafer, and the absence of thermal grooves in this image indicates
there are no grain boundaries over the entire area. (See Supplementary
Fig. S3 for EBSD maps of the rare grain boundaries that remain after
annealing.) The dark spots in Fig. \ref{Optical}\textbf{g} are areas
of dewetting, likely due to microscopic particles introduced during
transfer between deposition and annealing steps. Examination of such
films with the naked eye shows the Cu grains are typically $\gtrsim\unit[1]{cm}$
across. We have observed such giant grain growth on several 50~mm
wafers with $T_{\textrm{d}}=\unit[80]{^{\circ}C}$.

\begin{figure}[!h]
\centering{}\includegraphics[scale=0.4]{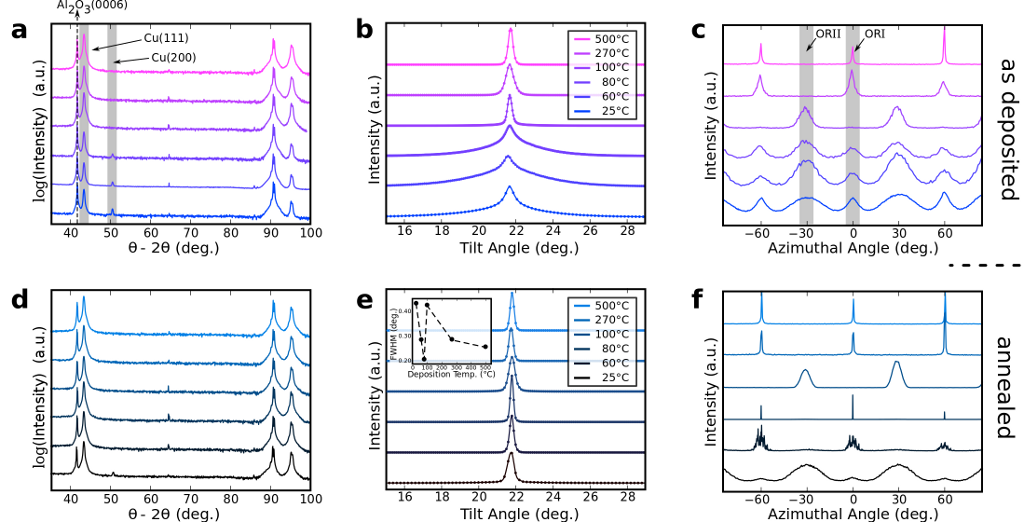}\caption{\textbf{X-ray diffraction results. a}-\textbf{c}, As-deposited films.
\textbf{d}-\textbf{f}, Annealed films. Legends in the center panels,
\textbf{b} and \textbf{e}, show the deposition temperature $T_{\textrm{d}}$.
The $\theta-2\theta$ scans in \textbf{a} and \textbf{d} show the
Cu film is predominately (111), with a small (100) component at lower
$T_{\textrm{d}}$ that decreases with annealing. The rocking curves
in \textbf{b} and \textbf{e} show the misalignment of <111> decreases
with annealing. The inset of\emph{ }\textbf{e} shows the width of
the rocking curve after annealing has a sharp minimum at $T_{\textrm{d}}=\unit[80]{^{\circ}C}$.
The azimuthal scans in \textbf{c} show the complex dependence of as-deposited
OR with $T_{\textrm{d}}$, while those in \textbf{f} show the dramatic
decrease in orientational disorder for the $T_{\textrm{d}}=\unit[80]{^{\circ}C}$
film.}
\label{XRD}
\end{figure}

It is evident from Fig. \ref{Optical} that giant grain growth only
occurs for a narrow range of $T_{\textrm{d}}$ near $\unit[80]{^{\circ}C}$.
In order to understand the conditions that lead to giant grain growth,
we have examined the structure of the Cu films at various $T_{\textrm{d}}$
using XRD before and after annealing. The XRD results for the as-deposited
films, Figs. \ref{XRD}\textbf{a}-\textbf{c}, reveal a sudden, qualitative
change in film properties as $T_{\textrm{d}}$ increases. For $T_{\textrm{d}}\leq\unit[80]{^{\circ}C}$,
we observe: (1) a small (100) component to the film texture (Fig.
\ref{XRD}\textbf{a}, note log scale), (2) broad tails in the rocking
curve, which measures misalignment of the <111> vector (Fig. \ref{XRD}\textbf{b}),
and (3) peaks corresponding to both OR~I and OR~II in the azimuthal
scan (Fig. \ref{XRD}\textbf{c}). The peaks every $60^{\circ}$ instead
of $120^{\circ}$ in the azimuthal scans indicate both OR~I and OR~II
consist of twins that are related by an in-plane rotation of $60^{\circ}$.
This twinning, which is expected because the OR between a (111) cubic
film and a hexagonal substrate is equivalent for a $60^{\circ}$ in-plane
rotation, has limited the crystallinity of Cu(111) in several graphene
CVD studies \cite{Miller:2012fk,Reddy:2011kx,Ishihara:2011kx,Hu:2012vn}.
For $T_{\textrm{d}}\geq\unit[100]{^{\circ}C}$, the film texture is
purely (111), the rocking curve has no tails, and the azimuthal scan
shows a single OR. The changes in these three features point to a
sudden transition to improved epitaxy between $T_{\textrm{d}}=\unit[80]{^{\circ}C}$
and $T_{\textrm{d}}=\unit[100]{^{\circ}C}$. Surprisingly, it is just
\emph{below} this transition where giant grain growth is most pronounced,
i.e., better as-deposited epitaxy does not necessarily promote giant
grain growth upon annealing.

The XRD results for the annealed films, Figs. \ref{XRD}\textbf{d}-\textbf{f},
show that annealing decreases polycrystallinity for all films: the
(100) component is reduced, the rocking curve tails for $T_{\textrm{d}}\leq\unit[80]{^{\circ}C}$
are eliminated, and the azimuthal scans show a clear preference for
one OR in all cases. The width of the rocking curve peak (inset of
Fig. \ref{XRD}\textbf{e}) shows a sharp minimum at the value of $T_{\textrm{d}}$
for which giant grain growth is most pronounced. The azimuthal scan
for the $T_{\textrm{d}}=\unit[80]{^{\circ}C}$ film shows a complete
absence of OR~II grains, with sharp peaks at the OR~I directions
whose width matches the instrumental limit set by beam divergence
(see Methods). The azimuthal scan for the $T_{\textrm{d}}=\unit[60]{^{\circ}C}$
film shows clusters of sharp peaks, indicating the presence of a few
large grains with discrete orientations near the nominal OR~I direction.

\begin{figure}[!h]
\centering{}\includegraphics[scale=0.4]{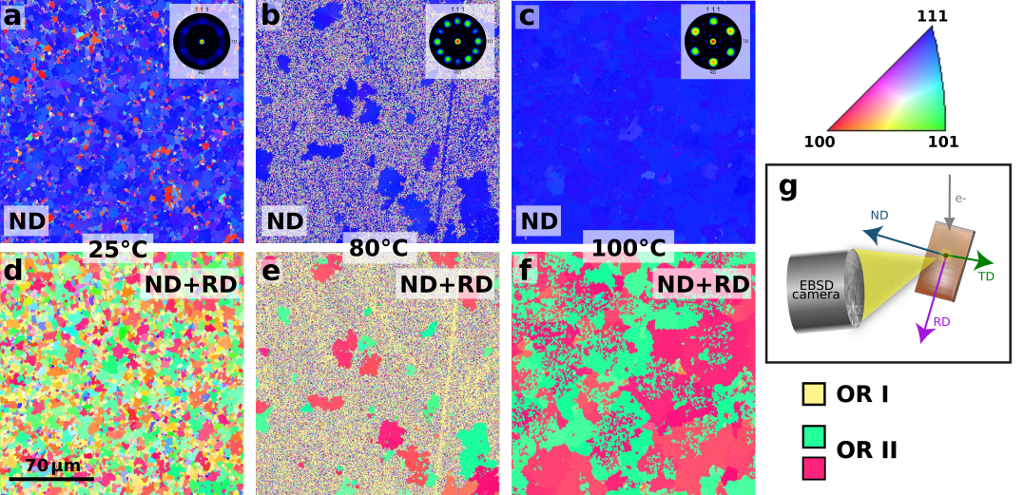}\caption{\textbf{EBSD texture maps of Cu orientation in as-deposited films.}
\textbf{a},\textbf{b},\textbf{c}, Maps using the normal direction
(ND) for $T_{\textrm{d}}$ = $\unit[25]{^{\circ}C}$, $\unit[80]{^{\circ}C}$
and $\unit[100]{^{\circ}C}$, respectively. Insets show the (111)
pole figure for each map. The color scale for these maps is shown
to the right of the images. \textbf{d},\textbf{e},\textbf{f}, Maps
using both normal and and in-plane components (ND+RD). Colors for
OR~I and OR~II are shown to the right of the images. The scale bar
in \textbf{d} applies to all images in this figure. \textbf{g}, EBSD
geometry showing ND, RD, and TD directions.}
\label{EBSD1}
\end{figure}

The XRD results point to a specific polycrystalline state of the as-deposited
film that favors giant grain growth. We have used EBSD to study the
microstructure of the Cu grains in this polycrystalline state. Texture
maps for $T_{\textrm{d}}$ = $\unit[25]{^{\circ}C}$, $\unit[80]{^{\circ}C}$
and $\unit[100]{^{\circ}C}$ are shown in Fig. \ref{EBSD1}\textbf{a}-\textbf{f}.
The geometry of the EBSD measurement, with definitions of reference,
transverse and normal directions (RD, TD, and ND), is shown in Fig.
\ref{EBSD1}\textbf{g}. Figures \ref{EBSD1}\textbf{a}-\textbf{c}
show ND maps, which indicate the out-of-plane direction according
to the color scale defined at the right side of the figure, with a
(111) pole figure as an inset to each map. Figures \ref{EBSD1}\textbf{d}-\textbf{f}
show ND+RD maps where the in-plane component highlights the distinction
between OR~I and OR~II, as shown by the colored squares at the right
side of the figure. (For this choice of map, the OR~II twins appear
as two different colors but both OR~I twins appear as a single color.)
These maps show a microstucture that is consistent with the XRD results
for the corresponding $T_{\textrm{d}}$ in Figs. \ref{XRD}\textbf{a}-\textbf{c}.
For $T_{\textrm{d}}=\unit[25]{^{\circ}C}$, the as-deposited film
is mostly (111) with a small (100) component that appears as a few
red grains in the ND map in Fig. \ref{EBSD1}\textbf{a}. The multiple
shades of blue in the ND map indicate many grains are somewhat misaligned
from (111), which is consistent with the broad XRD rocking curve tails.
Both the pole figure and the ND+RD map (Fig. \ref{EBSD1}\textbf{d})
show grains distributed broadly near OR~I and OR~II in roughly equal
proportion. For $T_{\textrm{d}}=\unit[100]{^{\circ}C}$, only grains
close to (111) appear in the ND map (Fig. \ref{EBSD1}\textbf{c}),
and the ND+RD map shows only OR~II grains (Fig. \ref{EBSD1}\textbf{f}).
For $T_{\textrm{d}}=\unit[80]{^{\circ}C}$, which favors giant grain
growth upon annealing, the as-deposited film is marked by a nonuniform
distribution of grain sizes: the ND map in Fig. \ref{EBSD1}\textbf{b}
shows several (111) grains that are much larger than their neighbors,
and the ND+RD map in Fig. \ref{EBSD1}\textbf{e} shows that these
large grains are exclusively OR~II. The pole figure has 12 spots
due to twinning in both the OR~II grains (pink and green in Fig.
\ref{EBSD1}\textbf{e}) and in the smaller OR~I grains (yellow in
Fig. \ref{EBSD1}\textbf{e}). The variation of colors for OR~II grains
is due to in-plane misalignment about the nominal epitaxial direction
of up to several degrees. This is consistent with the broad OR~II
peaks in the XRD azimuthal scans in Fig. \ref{XRD}\textbf{c}.

Although the large grain size for OR~II in Fig. \ref{EBSD1}\textbf{e}
indicates OR~II is favored over OR~I during deposition, the XRD
data of Fig. \ref{XRD}\textbf{f} show the OR~II grains are completely
converted into OR~I grains during annealing. This suggests the energy
difference between OR~I and OR~II is small and changes sign as temperature
is increased. Indeed, previous studies point to the same conclusion:
\emph{ab initio}, zero temperature calculations \cite{Hashibon:2005uq}
found OR~I and OR~II differ by 3\%, with OR~II favored, whereas
measurements of the shape of individual Cu particles created by solid-state
dewetting at \textbf{$\unit[980]{^{\circ}C}$} indicate a 4\% difference
with OR~I favored \cite{Curiotto:2011zr}. The different in-plane
strain expected for the two ORs, tensile for OR~I and compressive
for OR~II \cite{Hashibon:2005uq}, is qualitatively consistent with
this finding: since Cu expands more rapidly than $\alpha$-Al\textsubscript{2}O\textsubscript{3}(0001)
as temperature increases, the strain energy of an OR under tensile
strain will decrease and that of an OR under compressive strain will
increase.

\begin{figure}[!h]
\centering{}\includegraphics[scale=0.4]{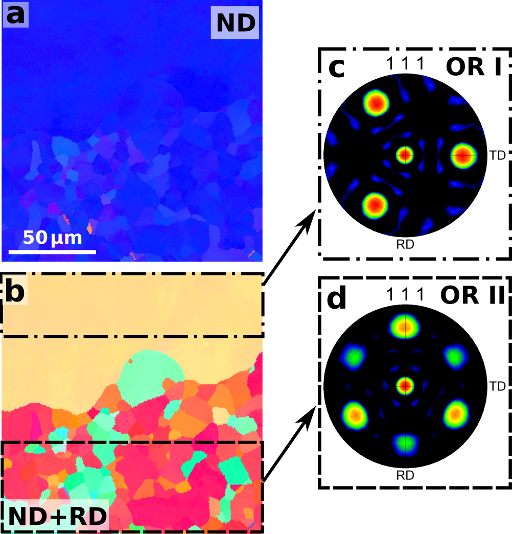}\caption{\textbf{EBSD maps of boundary between a giant grain and the neighboring
matrix.} \textbf{a}, Map using the normal direction (ND). \textbf{b},
Map of the same region using both normal and and in-plane components
(ND+RD). Color scales for both maps are the same as in Fig. \ref{EBSD1}.
\textbf{c}, Pole figure for the upper region of \textbf{b} shows untwinned
OR~I. \textbf{d}, Pole figure for the lower region of \textbf{b}
shows twinned OR~II.}
\label{BoundaryRegion}
\end{figure}

To examine the intermediate stages of giant grain growth, we used
shorter annealing times and lower annealing temperatures. For a $T_{\textrm{d}}=\unit[80]{^{\circ}C}$
film annealed for $\thicksim\unit[1]{min}$ at \textbf{$\unit[750]{^{\circ}C}$},
we found exclusively OR~II grains (Supplementary Fig. S4). After
annealing for 20~min at \textbf{$\unit[800]{^{\circ}C}$}, we found
individual OR~I grains starting to consume the surrounding matrix
of OR~II grains (Supplementary Fig. S2). Figure \ref{BoundaryRegion}
shows the boundary between a giant OR~I grain and the matrix for
a $T_{\textrm{d}}=\unit[80]{^{\circ}C}$ film annealed at $T_{\textrm{a}}=\unit[900]{^{\circ}C}$
for 40~min, i.e., as grain growth is nearing completion. Separate
pole figures for the upper and lower regions, Figs. \ref{BoundaryRegion}\textbf{c},\textbf{d},
show that the upper region is a single OR~I grain (no twinning) whereas
the matrix contains both OR~II twins. It is clear that the process
of giant grain growth involves OR~I grains consuming OR~II grains
upon annealing. In order for this to occur, the as-deposited film
must contain a fraction of both orientation relationships, because
if there are no OR~I seed grains present (as is the case with $T_{\textrm{d}}=\unit[100]{^{\circ}C}$)
or if the film is already exclusively OR~I ($T_{\textrm{d}}\geq\unit[270]{^{\circ}C}$),
the film will stagnate with the as-deposited orientation relationship,
creating thermal grooves between twins and eventually severe dewetting
(Supplementary Fig. S1). Deposition at $T_{\textrm{d}}=\unit[80]{^{\circ}C}$
provides a mix of grain orientations that promotes growth of giant
OR~I grains before significant dewetting occurs, thus stabilizing
the Cu film and allowing it to survive graphene CVD conditions. Annealing
of a $T_{\textrm{d}}=\unit[80]{^{\circ}C}$ film deposited through
a shadow mask to create isolated $\unit[100]{\text{\textmu m}}$ Cu
islands showed the density of OR~I seed grains is only about one
per mm\textsuperscript{2} (Supplementary Fig. S6). 

\begin{figure}[!th]
\centering{}\includegraphics[scale=0.4]{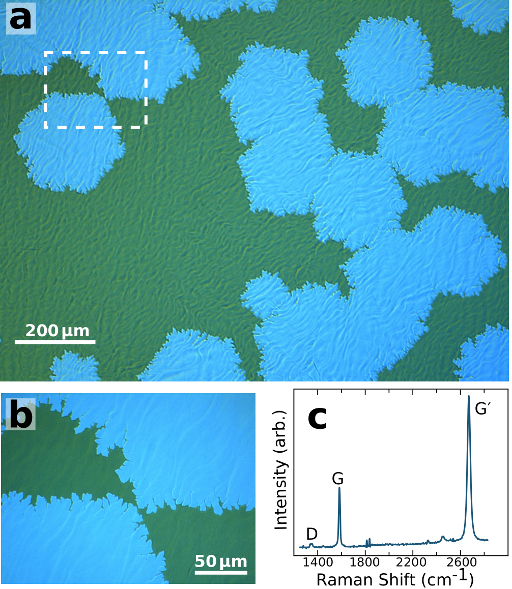}\caption{\textbf{Graphene growth on giant-grain Cu(111) films.} \textbf{a},
Optical image using differential interference contrast of graphene
on Cu after oxidation in air at $\unit[180]{^{\circ}C}$. Oxidized
Cu appears darker than the unoxidized regions covered by graphene.
\textbf{b}, Higher magnification image of the region of \textbf{a}
in the dashed box. \textbf{c}, Raman spectrum of graphene after transfer
to SiO\textsubscript{2} substrate.}
\label{Graphene}
\end{figure}

Graphene grown by CVD (see Methods) is shown in Fig. \ref{Graphene}.
We have optimized the growth parameters for large graphene domains
and growth on thin films. Notably, we use an Ar overpressure of $\unit[5300]{Pa}$
($\unit[40]{Torr}$) to suppress Cu sublimation and maintain thin
film stability. The image in Fig. \ref{Graphene}\textbf{a} was taken
after stopping growth before a complete graphene layer formed. The
sample was baked in air at $\unit[180]{^{\circ}C}$ for 5 min to oxidize
the bare Cu regions and thus provide optical contrast with unoxidized
regions covered by graphene. The graphene islands are roughly 6-fold
symmetric, as expected for a Cu(111) surface, and are $\approx\unit[200]{\mu m}$
across. Supplementary Fig. S7 contains additional images of graphene
grown on giant Cu grains. The graphene domains appear to be aligned,
suggesting epitaxy and consistent with other results \cite{Hu:2012vn}.
Since the Cu is epitaxial with the sapphire substrate, this enables
alignment of macroscopic features such as sample edges with zig-zag
and armchair directions in graphene nanostructures. Figure \ref{Graphene}\textbf{b}
shows the dendritic nature of the island perimeter. This type of growth
is similar to graphene growth at low pressures where the H\textsubscript{2}/CH\textsubscript{4}
ratio is close to 1. Figure \ref{Graphene}\textbf{c} is a Raman spectrum
of the graphene after transfering a continuous graphene film to a
300~nm oxidized Si substrate using PMMA and thermal release tape.
The I\textsubscript{G'}/I\textsubscript{G} peak ratio is 2.3, with
Lorentzian FWHM peak widths of 12~cm\textsuperscript{-1} and 27~cm\textsuperscript{-1}
for the G and G' peaks, respectively, and the D peak is small. All
these features indicate a high quality, monolayer graphene film.

We have demonstrated a route to single crystal Cu(111) films over
centimeter length scales based on dramatic secondary grain growth
of a favored orientation of Cu on $\alpha$-Al\textsubscript{2}O\textsubscript{3}(0001).
Our XRD and EBSD results show in detail the particular as-deposited
grain structure that promotes giant grain growth upon annealing. Although
this work has focused on Cu films thick enough to survive graphene
CVD conditions, we expect similar phenomena to occur at lower temperatures
for thinner films, based on standard grain growth models. Furthermore,
materials other than Cu should also be suitable for giant grain growth
if the required as-deposited grain structure can be obtained. A promising
candidate is Al on $\alpha$-Al\textsubscript{2}O\textsubscript{3}(0001),
since films having a mixture of OR~I and OR~II have already been
demonstrated \cite{Dehm:2005fk}.

\section*{{\normalsize Methods}}

\paragraph*{\textbf{\emph{Cu film deposition and annealing.}}}

Wafers of crystalline $\alpha$-Al\textsubscript{2}O\textsubscript{3}(0001),
50~mm in diameter, were annealed at $\unit[1100]{^{\circ}C}$ in
O\textsubscript{2} at atmospheric pressure for 24~h to remove scratches
due to polishing and give atomically flat terraces. Without further
processing, wafers were mounted on a resistively heated Cu puck and
placed in a vacuum sputter deposition system. Films of 450~nm nominal
thickness were deposited by dc magnetron sputtering from 76~mm targets
of 99.999\% pure Cu at a rate of $\approx\unit[1]{nm/s}$. Actual
film thickness measured using a profilometer ranged from 430~nm to
500~nm across the wafer, but this variation had no discernable effect
on the results. Films with $T_{\textrm{d}}<\unit[500]{^{\circ}C}$
were deposited in a load-locked, turbopumped chamber (base pressure
$\unit[2*10^{-7}]{Pa}$) with a target-to-substrate distance of $\unit[8]{cm}$
and a sputtering power of $\unit[100]{W}$ in $\unit[0.17]{Pa}$ $(1.2\unit[5]{mTorr})$
of Ar. Films deposited at $T_{\textrm{d}}\geq\unit[500]{^{\circ}C}$
were deposited in a cryopumped bell-jar-style chamber (base pressure
$\unit[2*10^{-5}]{Pa}$) with a target-to-substrate distance of $\unit[10]{cm}$
and a sputtering power of $\unit[200]{W}$ in $\unit[0.67]{Pa}$ $(\unit[5]{mTorr})$
of Ar. The deposition temperature $T_{\textrm{d}}$ reported here
was measured using a thermometer in contact with the sample puck.
After deposition, each wafer was coated with PMMA and diced into $\unit[5]{mm}\times\unit[6]{mm}$
chips. The chips were stripped of PMMA and cleaned by ultrasonic agitation
in acetone and isopropanol before characterization and annealing.
Annealing was performed in a hot-wall tube furnace with a base pressure
of $\unit[0.67]{Pa}$ $(\unit[5]{mTorr})$ at temperatures ranging
from $\unit[950]{^{\circ}C}$ to $\unit[1050]{^{\circ}C}$ for a duration
of 40~min. The temperature of the furnace was measured using a thermocouple
mounted just outside the quartz tube. To suppress Cu sublimation,
we maintained a total pressure of $\unit[5300]{Pa}$ $(\unit[40]{Torr})$
while flowing 500~sccm 99.999\% pure Ar and 11 sccm of 0.2\% H\textsubscript{2}
in Ar, yielding a H\textsubscript{2} partial pressure of $\unit[0.23]{Pa}$
$(\unit[1.7]{mTorr})$.

\paragraph*{\textbf{\emph{XRD and EBSD.}}}

X-ray diffraction was performed using a Cu K\textgreek{a} source and
a 4-circle goniometer with an instrument resolution of $0.0001^{\circ}$.
All measurements were performed using parallel beam optics with a
maximum beam divergence of $0.15^{\circ}$, which sets the minimum
rocking curve linewidth achievable in the experiment. A powder Si
sample was used to correct any offsets in the $2\theta$ angle. Symmetric
$\theta-2\theta$ scans were used to determine the out-of-plane crystalline
axes. Once the (111) reflection was found, the tilt angle was scanned
to generate the (111) rocking curve. In-plane azimuthal scans of the
(220) reflection were taken with the grazing incidence angle adjusted
to the value of maximum intensity (typically about $0.5^{\circ}$
above the critical angle). At this angle, the entire $\unit[5]{mm}\textrm{ x }\unit[6]{mm}$
sample is illuminated by the x-rays. Once the (220) peak was located,
an azimuthal scan about the sample normal axis was taken to determine
the in-plane angular distribution of the {[}220{]} axes.

For EBSD measurements, all films were oriented with \textcolor{black}{$[11\bar{2}0]_{\textrm{Al}{}_{2}\textrm{O}_{3}}$}
parallel to the RD direction. The SEM accelerating voltage was 20
kV, the chip was tilted $70^{\circ}$ towards the detector, and diffraction
patterns were collected with 4 $\times$ 4 binning over a hexagonal
array of \ensuremath{\approx} 500 nm pixels.

\paragraph*{\textbf{\emph{Graphene growth, transfer, and characterization.}}}

Before graphene growth, the Cu was annealed at $T_{\textrm{a}}=\unit[950]{^{\circ}C}$
for 1~h under 69.5 sccm of 0.2\% H\textsubscript{2} in Ar to form
large grains that are less susceptible to dewetting at higher temperatures.
The flow of 0.2\% H\textsubscript{2} in Ar was then decreased to
11 sccm, giving a H\textsubscript{2} partial pressure of $\unit[0.23]{Pa}$
$(\unit[1.7]{mTorr})$, and the temperature was ramped to $\unit[1055]{^{\circ}C}$
and stabilized for 5~min. Growth was initiated by introducing 14~sccm
of 0.2\% CH\textsubscript{4} in Ar, yielding a CH\textsubscript{4}
partial pressure of $\unit[0.28]{Pa}$ $(\unit[2.1]{mTorr})$. Growth
conditions were maintained for 40~min to achieve complete monolayer
coverage over an entire chip or wafer. Partial growths were achieved
by reducing the growth time. All gases were filtered at the mass flow
controller inlets to remove H\textsubscript{2}O and O\textsubscript{2}
contaminants to $\unit[\sim1]{ppb}$. For transfer of graphene to
oxidized Si, PMMA was spun onto the surface and baked at $\unit[150]{^{\circ}C}$,
and thermal release tape was applied to the PMMA. Chips were floated
on the surface of an ammonium persulfate bath overnight to etch away
the Cu. The Al\textsubscript{2}O\textsubscript{3} substrate fell
away, leaving the graphene/PMMA/tape layer on the liquid surface,
and this was then rinsed in DI water and placed on a $\unit[9]{mm}\textrm{ x }\unit[9]{mm}$
Si chip with $\unit[300]{nm}$ of SiO\textsubscript{2}. The Si chip
was heated to $\unit[150]{^{\circ}C}$ on a hot plate to release the
tape, and then to $\unit[180]{^{\circ}C}$ to soften the PMMA and
promote adhesion of graphene to the SiO\textsubscript{2} \cite{Suk:2011vn}.
To remove the PMMA layer, the chips were submerged in glacial acetic
acid for several hours at room temperature. Raman spectroscopy was
performed using an excitation wavelength of 532~nm.

\bibliographystyle{naturemag}

\section*{{\normalsize Acknowledgements}}

The authors thank Carl Thompson for helpful discussions of grain growth
phenomena. Optical microscope images were taken in NIST's Precision
Imaging Facility.

\section*{{\normalsize Author contributions}}

D.L.M. and M.W.K. designed the experiments, built custom equipment,
and performed film deposition. J.M.S. performed XRD measurements and
data analysis. K.P.R., R.R.K., and D.L.M. performed EBSD measurements
and analysis. D.L.M. and K.M.D. performed annealing, optical microscopy,
graphene growth, and graphene transfer. D.L.M. performed Raman spectroscopy.
M.W.K. and D.L.M. wrote the manuscript. All authors discussed the
data and commented on the manuscript.

\section*{{\normalsize Additional information}}

The authors declare no competing financial interests. Supplementary
information accompanies this paper on www.nature.com/naturematerials.
Reprints and permissions information is available online at http://npg.nature.com/reprintsandpermissions.
Correspondence and requests for materials should be addressed to M.W.K.
\end{document}